\documentclass[12pt]{article}
\usepackage{amsfonts,amsmath,amssymb,mathrsfs}

\title{A Kolmogorov Extension Theorem for POVMs\footnote{The main proof in this article was first formulated in my habilitation thesis \cite{Tum07b}.}}
\author{
Roderich Tumulka\footnote{Department of Mathematics,
     Rutgers University, 110 Frelinghuysen Road, 
     Piscataway, NJ 08854-8019,USA. E-mail:
     tumulka@math.rutgers.edu}
}
\date{October 18, 2006}

\newcommand{\Hilbert}{\mathscr{H}}

\newcommand{\be}{\begin{equation}}
\newcommand{\ee}{\end{equation}}

\newcommand{\PPP}{\mathbb{P}}
\newcommand{\RRR}{\mathbb{R}}
\newcommand{\CCC}{\mathbb{C}}

\newcommand{\NNN}{\mathbb{N}}
\newcommand{\scp}[2]{\langle #1|#2 \rangle}

\newcommand{\tr}{\mathrm{tr}}

\newcommand{\Bdd}{\mathscr{B}}
\newcommand{\salg}{\mathcal{A}}
\newcommand{\povm}{G}
\newcommand{\Borel}{\mathcal{B}}

\newtheorem{thm}{Theorem}

\newtheorem{cor}{Corollary}
\newtheorem{defn}{Definition}

\newenvironment{proofthm}[1]{\noindent\textit{Proof of Theorem~\ref{#1}. }}{\hfill$\square$\bigskip }
\newcounter{ex}

\begin{document}
\maketitle
\begin{abstract}
We prove a theorem about positive-operator-valued measures (POVMs) that is an analog of the Kolmogorov extension theorem, a standard theorem of probability theory. According to our theorem, if a sequence of POVMs $\povm_n$ on $\RRR^n$ satisfies the consistency (or projectivity) condition $\povm_{n+1}(A\times \RRR) = \povm_n(A)$ then there is a POVM $\povm$ on the space $\RRR^\NNN$ of infinite sequences that has $\povm_n$ as its marginal for the first $n$ entries of the sequence. We also describe an application in quantum theory.
\medskip

\end{abstract}

\section{Introduction}

A relevant mathematical concept for quantum physics is that of POVM (positive-operator-valued measure). It forms the natural generalization of the concept of observable represented by a self-adjoint operator. One can say that all probability measures that arise in quantum physics are of the form
\be\label{povm}
\PPP(\cdot) = \tr\bigl(\rho\, \povm(\cdot)\bigr)\,,
\ee
where $\rho$ is a density matrix in the appropriate Hilbert space $\Hilbert$ and $\povm$ is a POVM. We recall the definition of POVM in Section~\ref{sec:defpovm}.

The Kolmogorov extension theorem is a standard theorem of probability and measure theory concerning probability measures \cite{Kal97}. In its simplest version (due to Daniell) \cite[Theorem 5.14]{Kal97} it asserts that if for every $n\in\NNN$, $\mu_n$ is a probability measure on $\RRR^n$ (with its Borel $\sigma$-algebra $\Borel^{\otimes n}$) such that the so-called \emph{consistency (or projectivity) condition}
\be\label{consistency}
\mu_{n+1}(A\times\RRR) = \mu_n(A)\quad \forall A\in \Borel^{\otimes n}
\ee
is satisfied then there exists a unique probability measure $\mu$ on the space $\RRR^\NNN$ of all real sequences (with the $\sigma$-algebra $\Borel^{\otimes \NNN}$ generated by the cylinder sets, i.e., sets depending only on finitely many members of the sequence) such that $\mu_n$ is its marginal distribution of the first $n$ components. Note that \eqref{consistency} is a necessary condition for $\mu_n(\cdot)$ being a marginal of some probability measure on $(\RRR^\NNN,\Borel^{\otimes \NNN})$. The more refined versions of the Kolmogorov extension theorem we describe in Section~\ref{sec:Kol}

In our theorem we replace the probability measures with POVMs. The proof utilizes the Kolmogorov extension theorem, but has to take care of the quadratic dependence of the probabilities on $\psi$. In the remainder of Section 1, we provide more detail about the Kolmogorov extension theorem and the concept of POVM. In Section 2 we formulate and prove our theorem. In Section 3 we describe an application in quantum physics.

\subsection{The Kolmogorov Extension Theorem}
\label{sec:Kol}

Above we formulated the simplest version of the Kolmogorov extension theorem. The statement remains true if $\RRR$ is replaced by any Borel space. Recall that a Borel space is a measurable space $(M,\salg)$ that is isomorphic (in the category of measurable spaces) to a measurable subset of the real line. Any Polish space (i.e., complete separable metric space) with its Borel $\sigma$-algebra is a Borel space; in particular, $\RRR$, $\RRR^n$, separable Hilbert spaces and separable manifolds are all Borel spaces.

The statement also remains true if, instead of $M^n$ and $M^\NNN$, one considers $M_1 \times \cdots \times M_n$ and $\prod_{n\in\NNN} M_n$, provided every $(M_n,\salg_n)$ is a Borel space, with the $\sigma$-algebras $\salg_1 \otimes \cdots \otimes \salg_n$ and $\bigotimes_{n\in\NNN} \salg_n$ (the $\sigma$-algebra generated by the cylinder sets). 

The statement still remains true if $\NNN$ is replaced by any (possibly uncountable) set $T$. The name ``Kolmogorov extension theorem'' often refers to this version in particular. Then it is to be formulated as follows. Let $\hat T$ be the collection of all finite subsets of $T$. Suppose we are given a family $\{(M_t,\salg_t): t\in T\}$ of Borel spaces and a family $\{\mu_K: K \in \hat T\}$ of probability measures $\mu_K$ on $(M^K,\salg^K) := (\prod_{t\in K}M_t, \bigotimes_{t\in K} \salg_t)$ satisfying the consistency condition
\be\label{Kolcon}
K \subseteq K' \:\Longrightarrow\: \forall A\in\salg^K: \mu_{K'}\Bigl(A\times\prod_{t\in K'\setminus K}M_t\Bigr) = \mu_K(A)\,.
\ee
Then there is a unique probability measure $\mu$ on $(M^T,\salg^T) := (\prod_{t\in T} M_t, \bigotimes_{t\in T} \salg_t)$ (where the $\sigma$-algebra is again generated by the cylinder sets, i.e., sets depending only on finitely many $t$'s) such that $\mu_K$ is the marginal of $\mu$ on $M^K$.

\subsection{The Concept of POVM}
\label{sec:defpovm}

Let $\Hilbert$ be a complex Hilbert space (not necessarily separable) and $\Bdd(\Hilbert)$ the space of bounded operators on $\Hilbert$.
We recall the following (standard) definition.

\begin{defn}
A \textbf{POVM} (positive operator valued measure) on the measurable space $(\Omega,\salg)$ acting on $\Hilbert$ is a mapping $\povm:\salg \to \Bdd(\Hilbert)$ from a $\sigma$-algebra $\salg$ on the set $\Omega$ such that
\begin{itemize}
\item[(i)] $\povm(\Omega) = I$, the identity operator, 
\item[(ii)] $\povm(A) \geq 0$ (i.e., $\povm(A)$ is a positive operator) for every $A \in \salg$, and
\item[(iii)] ($\sigma$-additivity) for any sequence of pairwise disjoint sets $A_1, A_2, \ldots \in \salg$
\begin{equation}\label{sigmaadditive}
    \povm\Bigl( \bigcup_{i=1}^\infty A_i \Bigr) = \sum_{i=1}^\infty \povm(A_i)\,,
\end{equation}
where the sum on the right hand side converges weakly, i.e., $\sum_i \scp{\psi}{\povm(A_i)\, \psi}$ converges for every $\psi\in\Hilbert$ to $\scp{\psi}{\povm(\cup_i A_i)\psi}$. 
\end{itemize}
\end{defn}

\medskip

If $\povm$ is a POVM on $(\Omega,\salg)$ and $\psi\in\Hilbert$ with $\|\psi\|=1$, then $A \mapsto \scp{\psi}{\povm(A)\, \psi}$ is a probability measure on $(\Omega,\salg)$.

\section{Theorem and Proof}

\begin{thm}\label{thm:Kol}
Let $\Hilbert$ be a Hilbert space, $T$ be an arbitrary index set, and for every $t\in T$ let $(M_t,\salg_t)$ be a Borel space. Let $\hat T$ denote the collection of all finite subsets of $T$. For every $K\in \hat T$ let $\povm_K$ be a POVM on $(M^K,\salg^K) := (\prod_{t\in K}M_t, \bigotimes_{t\in K} \salg_t)$ acting on $\Hilbert$. If the $\povm_K$ satisfy the consistency condition
\be\label{povmKolcon}
K \subseteq K' \:\Longrightarrow\: \forall A\in\salg^K: \povm_{K'}\Bigl(A\times\prod_{t\in K'\setminus K}M_t\Bigr) = \povm_K(A)
\ee
then there is a unique POVM $\povm$ on $(M^T,\salg^T) := (\prod_{t\in T} M_t, \bigotimes_{t\in T} \salg_t)$ such that $\povm_K$ is the marginal of $\povm$ on $M^K$.
Moreover, for every $\psi\in \Hilbert$ with $\|\psi\|=1$ there exists a unique probability measure $\mu^\psi$ on $(M^T,\salg^T)$ such that for all $K\in \hat T$ and all sets $A\in\salg^K$,
\be
\mu^\psi\Bigl(A\times \prod_{t\in T\setminus K}M_t\Bigr) = \scp{\psi}{\povm_K(A)\,\psi}\,,
\ee
and in fact $\mu^\psi(\cdot) = \scp{\psi}{\povm(\cdot)\,\psi}$.
\end{thm}

(To appreciate why the last sentence is not a reformulation of the previous ones but an independent statement, the reader should note that even if the POVM $\povm$ that extends all the $\povm_K$ is unique, we must exclude the possibility of a further measure $\tilde{\mu}^\psi$ that is \emph{not} given by a POVM but does extend the $\mu_K^\psi(\cdot) = \scp{\psi}{\povm_K(\cdot)\,\psi}$.)

As a special case of the theorem, we obtain the following statement for $T=\NNN$ and $(M_t,\salg_t) = (M,\salg)$.

\begin{cor}\label{cor:Dan}
Let $(M,\salg)$ be a Borel space and $\povm_n$, for every $n\in\NNN$, a POVM on $(M^n,\salg^{\otimes n})$. If the family $\{\povm_n\}_{n\in\NNN}$ satisfies the consistency property
\begin{equation}\label{FPOVMconsistent}
  \povm_{n+1}(A \times M) = \povm_n(A) \quad \forall A\in\salg^{\otimes n}
\end{equation}
then there exists a unique POVM $\povm$ on $(M^\NNN, \salg^{\otimes \NNN})$ 
such that for all $n\in\NNN$ and all sets $A\in\salg^{\otimes n}$,
\begin{equation}\label{FPOVMcyl}
  \povm_n(A) = \povm(A \times M^\NNN) \,.
\end{equation}
Moreover, for every $\psi\in \Hilbert$ with $\|\psi\|=1$ there exists a unique probability measure $\mu^\psi$ on $(M^\NNN,\salg^{\otimes\NNN})$ such that for all $n\in\NNN$ and all sets $A\in\salg^{\otimes n}$, $\mu^\psi(A\times M^\NNN) = \scp{\psi}{\povm_n(A)\,\psi}$, and in fact $\mu^\psi(\cdot) = \scp{\psi}{\povm(\cdot)\,\psi}$.
\end{cor}

\begin{proofthm}{thm:Kol}
For arbitrary $\psi\in\Hilbert$ with $\|\psi\|=1$, define the probability measure
\begin{equation}
\mu_K^\psi(\cdot) = \scp{\psi}{\povm_K(\cdot)\, \psi}
\end{equation}
on $(M^K,\salg^K)$.
Because of \eqref{povmKolcon}, \eqref{Kolcon} is fulfilled, so the family $(\mu_K^\psi)_{K\in \hat T}$ of probability measures is consistent, and by the Kolmogorov extension theorem there exists a unique measure $\mu^\psi$ on $(M^T, \salg^T)$ such that for all $K \in \hat T$ and all $A_K\in\salg^K$,
\be
\mu_K^\psi(A_K) = \mu^\psi\Bigl(A_K\times \prod_{t\in T\setminus K}M_t\Bigr)\,.
\ee

To see that $\mu^\psi(\cdot) = \scp{\psi}{\povm(\cdot)\, \psi}$ for some POVM $\povm(\cdot)$, we first define, for every $A \in \salg^T$ and every $\psi \in \Hilbert$, the complex number $\tilde{\mu}^\psi(A)$ by
\begin{equation}\label{tildemudef}
\tilde\mu^\psi (A):=\begin{cases} \|\psi\|^2 \, \mu^{\psi/\|\psi\|}(A) & \text{if }\psi\neq 0\\
0 & \text{if }\psi=0 \,,\end{cases}
\end{equation}
which allows us to use any $\psi \in \Hilbert$, also with $\|\psi\| \neq 1$.
Furthermore, we define, for every $\psi,\phi \in\Hilbert$, the complex number $\mu_{\psi,\phi}(A)$ by ``polarization'':
\begin{equation}\label{polarization}
  \mu_{\psi,\phi}(A) := \tilde\mu^{\frac{1}{2}\psi + \frac{1}{2} \phi}(A) -
  \tilde\mu^{\frac{1}{2}\psi - \frac{1}{2}\phi}(A) +
  i\tilde\mu^{\frac{1}{2}\psi-\frac{i}{2}\phi}(A) -
  i\tilde\mu^{\frac{1}{2}\psi + \frac{i}{2} \phi}(A)\,,
\end{equation}
The definitions \eqref{tildemudef} and \eqref{polarization} are so chosen that if $A$ is a cylinder set,
\begin{equation}\label{cylinderA}
A=A_K \times \prod_{t\in T\setminus K} M_t\,,
\end{equation}
then
\begin{equation}\label{musp}
  \tilde\mu^\psi(A) = \scp{\psi}{\povm_K(A_K) \, \psi} \quad\text{and}\quad
  \mu_{\psi,\phi}(A) = \scp{\psi}{\povm_K(A_K) \, \phi}\,.
\end{equation}
Note that $\mu_{\psi,\phi}$ is a complex measure on $(M^T,\salg^T)$ since, by \eqref{polarization}, it is a complex linear combination of finite measures. Since the cylinder sets form a $\cap$-stable generator of $\salg^T$, two complex measures on $\salg^T$ coincide as soon as they agree on the cylinder sets. Therefore, the three equations 
\begin{equation}
  \mu_{\psi+\psi',\phi}(A) = \mu_{\psi,\phi}(A) + \mu_{\psi',\phi}(A) 
  \quad \forall \psi,\psi',\phi \in \Hilbert\,,
\end{equation}
\begin{equation}
  \mu_{z\psi,\phi}(A) = z^* \, \mu_{\psi,\phi}(A)
  \quad \forall z\in \CCC \forall \psi,\phi \in \Hilbert\,,
\end{equation}
and
\begin{equation}
  \mu_{\phi,\psi}(A) = \mu_{\psi,\phi}(A)^* \quad \forall \psi,\phi \in \Hilbert \,,
\end{equation}
which hold for cylinder sets $A$ because of \eqref{musp}, hold for all $A \in \salg^T$. Thus, for every fixed $A\in\salg^T$, $(\psi,\phi) \mapsto \mu_{\psi,\phi}(A)$ is a Hermitian sesquilinear form on $\Hilbert$. Since
\be
\mu_{\psi,\psi}(A) = \tilde{\mu}^\psi(A) = \|\psi\|^2 \, \mu^{\psi/\|\psi\|}(A) \leq \|\psi\|^2\,,
\ee
this sesquilinear form is bounded and has, in fact, norm $\leq 1$. Therefore, by the Riesz lemma \cite[p.~43]{RS72} there is a bounded operator $\povm(A)$ such that
\begin{equation}\label{GAdef}
\mu_{\psi,\phi}(A) = \scp{\psi}{\povm(A) \, \phi}\,.
\end{equation}
The uniqueness of $\povm(A)$ follows from the fact that if $\povm'(A) \neq \povm(A)$ then there is $\psi\in\Hilbert$ with $\|\psi\|=1$ such that $\scp{\psi}{\povm'(A)\, \psi} \neq \scp{\psi}{\povm(A)\,\psi}$. We now have that $\tilde\mu^\psi(A) = \mu_{\psi,\psi}(A) = \scp{\psi}{\povm(A)\, \psi}$. 

We check that $\povm$ is a POVM: $\povm(A)$ is positive since $\mu_{\psi,\psi}(A) = \tilde\mu^\psi(A) \geq 0$ for every $\psi$; $\povm(\cdot)$ is $\sigma$-additive in the weak sense because $\mu_{\psi,\phi}(\cdot)$ is $\sigma$-additive; $\povm(\cdot)$ restricted to the cylinder sets $A_k \times \prod_{t\in T\setminus K}M_t$ is $\povm_K(\cdot)$, and thus $\povm(M^T) = I$.

The uniqueness statement for the extension $\povm(\cdot)$ of all $\povm_K(\cdot)$'s follows from the uniqueness of $\povm(A)$ satisfying \eqref{GAdef}, together with the fact that for every fixed $\psi$ the measure $\mu^\psi$ is unique.

It remains to check that $\povm_K$ is the appropriate marginal of $\povm$. Let $A$ be of the form \eqref{cylinderA} with $A_K\in\salg^K$. For all $\psi\in\Hilbert\setminus\{0\}$,
\be
\scp{\psi}{\povm(A) \, \psi}= \tilde\mu^\psi(A) = \|\psi\|^2 \, \mu^{\psi/\|\psi\|}(A) = \|\psi\|^2\, \mu_K^{\psi/\|\psi\|}(A_K) = \scp{\psi}{\povm_K(A_K)\,\psi}\,.
\ee
But a positive bounded operator $T$ is uniquely determined by the family of numbers $\scp{\psi}{T\,\psi}$, hence $\povm(A) = \povm_K(A_K)$. 
\end{proofthm}

To see that Corollary 1 is a special case of Theorem 1, we have to check that, instead of considering \emph{all} finite subsets $K$ of $\NNN$, it suffices to consider the sets $\{1,\ldots,n\}$. Indeed, if the consistency relation \eqref{FPOVMconsistent} holds then we can define
\be
\povm_K (A) = \povm_n\Bigr(A \times \prod_{t\in \{1\ldots n\} \setminus K} M_t\Bigl)\,,
\ee
where the right hand side does not depend on the choice of $n$, provided that $n\geq \max K$. Through this definition, we obtain a POVM $\povm_K$, and the family $\{\povm_K\}_{K \in \hat \NNN}$ satisfies the consistency condition \eqref{povmKolcon}.

\section{Application}

The application that has caused me to formulate Theorem~\ref{thm:Kol} and to develop its proof concerns the Ghirardi--Rimini--Weber (GRW) theory of spontaneous wave function collapse \cite{GRW86,Bell87}, a \emph{quantum theory without observers} that has been proposed as a precise version and explanation of quantum mechanics. In the GRW theory, the collapse of the wave function is an objective physical event governed by a stochastic law, whereas in conventional quantum mechanics the collapse is said to occur whenever an ``observer'' intervenes. The evolution of the wave function is a stochastic process in Hilbert space, replacing the deterministic unitary Schr\"odinger evolution. The proof of the rigorous existence of this process \cite{Tum07b}, and thus the proof that the GRW theory is mathematically well defined, is best done by means of Theorem~\ref{thm:Kol}. In more detail, since each collapse in the GRW theory is associated with a space-time point, the wave function process in Hilbert space can be equivalently translated into a point process in space-time, that is, a random sequence of space-time points or a random element of $M^\NNN$, where $M$ is the space-time manifold. Given the stochastic GRW law of wave function evolution, the joint distribution of the first $n$ points can be written down explicitly, and in fact is of the form $\mu_n(\cdot) = \scp{\psi}{\povm_n(\cdot)\,\psi}$, where $\psi\in\Hilbert$ is the initial state vector with $\|\psi\|=1$ and $\povm_n$ is a POVM on $M^n$ acting on $\Hilbert$.

\end{document}